\documentclass{basi}
\usepackage{rotating}
\usepackage{psfig}

\begin{document}
\title[Indian Astronomical Observatory] {Night sky at the Indian Astronomical Observatory during 2000$-$2008}
\author[C. S. Stalin et al]
{C. S. Stalin\thanks{e-mail:stalin@iiap.res.in}, M. Hegde, D. K. Sahu, 
P. S. Parihar, G. C. Anupama, 
\newauthor
B. C. Bhatt and T. P. Prabhu \\
Indian Institute of Astrophysics, Block II, Koramangala, Bangalore 560 034, India}
\maketitle

\begin{abstract}
This paper presents an analysis of the optical night
sky brightness and extinction coefficient measurements in $UBVRI$ at the Indian Astronomical
Observatory (IAO), Hanle, during the period
2003$-$2008. They are obtained from an analysis of CCD images acquired at
the 2 m Himalayan Chandra Telescope (HCT) at IAO. 
Night sky brightness was estimated using 210 HFOSC images
obtained on 47 nights and covering the declining phase of solar activity
cycle-23. The zenith corrected values of the moonless night sky brightness in 
mag arcsec$^{-2}$ are 22.14 $\pm$ 0.32 ($U$), 22.42 $\pm$ 0.30
($B$), 21.28 $\pm$ 0.20 ($V$), 20.54 $\pm$ 0.37 ($R$) and 18.86 $\pm$ 0.35 
($I$) band. This shows that IAO is a dark site for optical observations.
No clear dependency of sky brightness with solar activity (implied by 
the 10.7 cm solar flux) is found. Extinction values at IAO are derived
from an analysis of 1325 images over 58 nights. They are found to be
0.36 $\pm$ 0.07 in $U$-band, 0.21 $\pm$ 0.04 in $B$-band, 
0.12 $\pm$ 0.04 in $V$-band, 0.09 $\pm$ 0.04 in $R$-band and 
0.05 $\pm$ 0.03 in $I$-band.
On average, extinction during the summer months is slightly larger
than that during the winter months. 
This might be due to an increase of dust
in the atmosphere during the summer months. 
No clear evidence for a correlation between extinction in all bands
and the average night time wind speed is found. 
Also presented here is the low resolution moonless optical night sky
spectrum for IAO covering the wavelength range 3000 $-$ 9300 \AA. 
Features from O, OH,  N and Na are seen in the spectra.
 Hanle region thus has the required characteristics
of a good astronomical site in terms of night sky brightness and
extinction, and could be a natural candidate site
for any future large aperture Indian optical-infrared telescope(s).

\end{abstract}

\begin{keywords}
atmospheric effects, site testing 
\end{keywords}

\section{Introduction}
%Knowledge of night sky brightness and atmospheric extinction at a given
%astronomical site is fundamental to ground based optical and infrared
%observations. 
A good astronomical site is characterised by various atmospheric
conditions (which includes atmospheric transparency, seeing, meteorological 
parameters such as wind, snowfall, surface temperature, rainfall etc. and sky brightness)
and geographical conditions (such as local topography, seismicity,
source availability i.e, latitude etc.). Sites having
minimum cloud coverage, very low frequency of
snowfall/rainfall, low relative humidity, low nocturnal temperature
variation, high atmospheric transparency and low night
sky brightness are good for ground based optical and infrared observations. 
Two main characteristics of the night sky are the night sky brightness
and atmospheric extinction.

%\subsubsection{Night sky brightness}

Even in the absence of artificial light, the moonless night sky is 
not dark. This is because the atmosphere scatters into the sky,
light emitted by the following processes (cf. Krisciunas 1997), (i) zodiacal 
light (caused by sunlight scattered 
off interplanetary dust), (ii) faint unresolved stars and diffuse galactic 
light due to atomic processes within our galaxy,
(iii) diffuse extragalactic light (due to distant, faint unresolved
galaxies) and (iv) airglow and aurorae (produced by photochemical reactions
in the Earth's upper atmosphere). Of these, (i)$-$(iii) are extraterrestrial
in origin and thus independent of the site, whereas (iv) depends on 
the site and time of observation. These are the  natural processes 
which produce the night sky brightness in any astronomical site.  In 
addition to the above, night sky can be affected by light pollution 
due to scattering of street lights in the Earth's lower atmosphere. 
Humans do not have control on any
of the natural sources causing the brightness of the night sky, but do 
have a control on the  
brightness caused by artificial lights scattered on to the sky. It is thus
possible to maintain the night sky brightness at any observatory site to its
natural level by minimising light pollution in the immediate vicinity of the 
observatory.

%\subsubsection{Extinction}
Apart from the natural and artificial sources affecting the night sky,
the light coming from any celestial source being observed 
suffers from scattering by air molecules
and aerosols as well as absorption by water vapour and ozone while 
passing through the Earth's atmosphere. This leads to 
attenuation of their light and is
referred to as the atmospheric extinction. This depends on the constituents
of the atmosphere, the wavelength of the incoming light, and the altitude
of the site. Precise knowledge of the extinction coefficient of each site
is essential to compare observations of the same object taken from 
different locations of the globe. A good astronomical site needs low
extinction values. Apart from low extinction, its stability during a
night is also equally important. 
%Humans do not have control on any of the above
%parameters, except in the case of night sky brightness, where it is
%possible to maintain at the natural level by minimizing light pollution
%at the immediate vicinity of the Observatory.   

In this article we present 
the moonless night sky brightness and atmospheric
extinction in $UBVRI$ passbands at IAO, Hanle. IAO is located at
the Himalayan range in Northern India (longitude = 
78$^{\circ}$57$^{\prime}$51.2$^{\prime\prime}$ E, 
latitude = 32$^{\circ}$46$^{\prime}$46.5$^{\prime\prime}$ N and altitude
= 4467 m) and run by the Indian Institute of Astrophysics, Bangalore. This is 
a thinly populated, cold and dry desert region. 
The sky at IAO is thus not much affected by dust and light pollution due to
human activities.  A 2 m telescope, the Himalayan Chandra 
Telescope (HCT) is operational at IAO since May 2003. 
The data used in this study for night sky brightness span the
period 2003$-$2008 which correspond to a major part of the declining
phase of solar activity (which could affect night sky brightness)
cycle-23, while the data for extinction estimates span the 
period 2000$-$2008. Preliminary estimates of the night sky brightness and
extinction at IAO have been reported by Parihar et al. (2003) and
an analysis of the meteorological parameters at IAO is presented in Stalin et al. (2008). 
The structure 
of this paper is as follows. Section 2 
describes the data set used in this study, Section 3 presents
the analysis of extinction and Section 4 presents
an analysis of night sky brightness. A spectrum of the night sky
at IAO is presented in Section 5 and the results are summarized in the 
final section.

\section{Data}
No data were obtained specifically for studying the night sky at IAO.
Therefore, archives at IAO were searched for a data set which 
is as homogeneous as possible. Multiband imaging data from the
supernova monitoring program of IAO were thus collected from
the archives spanning the years 2003$-$2008. They were from 
science observations carried
out using HFOSC at the 2 m HCT. The CCD used in these observations was 
a 2k $\times$ 4k, with a pixel scale of 0.296$^{\prime\prime}$/pix giving 
a sky coverage of 10 $\times$ 10 arcmin$^2$.
A total of 210 images obtained over 47 nights were extracted 
from the IAO archives and used to estimate the night sky brightness. 
%For extinction measurements, Landolt star fields
%were observed over several nights alloted for 
%preventive maintenance and calibration during 2000$-$2008. 
%This leads us to gather 893 HFOSC frames over 34 nights observed
%over a large range of airmass from 1.01 to 2.40.
%From 2000$-$2002, a 1k $\times$ 1k CCD was used for the observations to 
%find extinction, thereby acquiring 432 frames over 22 nights. From
%Feb. 2003 onwards, HFOSC was used for the extinction measurements. Using 
%HFOSC, 893 frames were acquired over 36 nights. Thus, for the analysis
%of the extinction presented here, a total of 1325 image frames over 58 nights
%and covering a large range of airmass from 1.01 to 2.40 were used.
Photometric standard fields (Landolt 1992) observed over several nights
during 2000$-$2008 were used to estimate the site extinction.
A 1k $\times$ 1k CCD system was in use during 2000$-$2002, while the 
HFOSC was used since 2003 February. The data used for extinction estimates
thus comprise of a total of 1325 image frames in the $UBVRI$ bands obtained
over 58 nights, covering an airmass range of 1.01 $-$ 2.40.
Pre-processing of all the photometric data  as well as photometry have
been done using IRAF\footnote{IRAF is distributed by the National Optical
Astronomy Observatories, which is operated by the Association of 
Universities for Research in Astronomy Inc. under contract to the 
National Science Foundation}.

%To get the night sky spectrum at IAO, spectroscopic data taken for 
%SN 2004et on the night of 16 October 2004,  was extracted from the IAO archives.
The night sky spectrum at IAO has been extracted from the spectroscopic data of
supernova SN 2004et obtained on 2004 October 16.
The spectra were obtained with a 11$^{\prime}$ long and 1.92$^{\prime\prime}$ 
wide slit and two grisms; grism 7 and grism 8 covering the 
wavelength range from 3500$-$7000 \AA ~and 5200$-$9200 \AA ~respectively. The spectral
resolution is 8 \AA. Spectroscopic
data too, was bias subtracted, flux and wavelength calibrated using 
standard IRAF procedures.

\section{Extinction}

The Bouguer's linear formula for atmospheric extinction is

\begin{equation}
m(\lambda,z) = m_{o}(\lambda) + 1.086 k_{\lambda} secz
\end{equation}

where $m(\lambda,z)$ is the observed magnitude, $m_{o}(\lambda)$ is the 
magnitude above the Earth's atmosphere, $k_{\lambda}$ is the extinction
and secz is the airmass at zenith distance z.

The three sources of extinction in the
Earth's atmosphere that are important for ground based astronomical
photometry are (Hayes \& Latham 1975)

\begin{enumerate}
\item $A_{aer}$ = Aerosol scattering
\item $A_{Ray}$ = Rayleigh scattering by molecules
\item $A_{oz}$ = molecular absorption mainly by ozone
\end{enumerate}

The contribution of each of these parameters to extinction depends
on wavelength, whereas, Rayleigh and Aerosol scattering, apart
from wavelength, depend also on height and atmospheric
conditions at the site.

According to Hayes \& Latham (1975), Rayleigh scattering by air
molecules at an altitude $h$ is given by

\begin{equation}
A_{Ray}(\lambda,h) = 9.4977 \times 10^{-3} \left(\frac{1}{\lambda}\right)^4 
C^2 \times exp\left(\frac{-h}{7.996}\right)
\end{equation}

where $C = 0.23465 + \frac{1.076 \times 10^2}{146-(1/\lambda)^2} + \frac{0.93161}{41-(1/\lambda)^2}$

Here, $\lambda$ is the wavelength in microns and $h$ is the altitude in km.
Eq.2 assumes an atmospheric pressure of 760 torr at $h$ = 0 and a scale
height of 7.996 km. The 
largest uncertainty here is due to the deviation of local atmospheric
pressure from the assumed standard condition.

Molecular absorption by ozone and water vapour too contribute to the total
extinction at any site.  Ozone is concentrated at altitudes between 10 and 
35 km and hence its contribution to the extinction does not depend on the 
altitude of the observatory. However, it is a function of wavelength and 
occurs in selective bands centered at $\lambda\lambda$
3300 and 5750 \AA ~(Gutierrez-Moreno et al. 1982). On the other hand, 
extinction due to water vapour is difficult to estimate, because
the amount of water vapour above a site is variable. This is however, 
weak and centered only around a few select bands with neglible
contribution to broad band.
The extinction due to ozone is (cf. Bessel 1990; Kumar et al. 2000)

\begin{equation}
A_{oz} = 0.2775 C_{oz}(\lambda)
\end{equation}

where $C_{oz}(\lambda)$ is the ozone absorption coefficient and is given 
as 

\begin{equation}
C_{oz}(\lambda) = 3025 exp \left(-131 (\lambda - 0.26)\right) + 0.1375 exp \left(-188 (\lambda - 0.59)^2\right)
\end{equation}

Extinction due to aerosol scattering is highly variable. This is due
to particulates including mineral dust, salt particles and man made
pollutants and is expressed as

\begin{equation}
A_{aer}(\lambda,h) = A_{o}\lambda^{-\alpha}exp(-h/H)
\end{equation}

where $H$ is the density scale height for aerosols and $A_o$ is the total
optical thickness of atmospheric aerosols for $\lambda$ = 
1 $\mu m$, which depends on the total content of particles and on their 
efficiency for scattering and absorption and is taken 
to be 0.087 (Mohan et al. 1999; Bessel 1990). $\alpha$ is a parameter
which depends on the size of the aerosol grains. Following 
Hayes \& Latham (1975)
a value of $H$ = 1.5 km and $\alpha$ = 0.8 is considered in this work.
The largest uncertainty in $A_{aer}$ from Eq. 5,  
is due to the incomplete knowledge of the nature
of aerosols at the location of IAO.

The total extinction at any given wavelength is therefore a linear
combination of these three contributions and is given as

\begin{equation}
A_{\lambda} = A_{aer}(\lambda,h) + A_{Ray} (\lambda, h) + A_{oz}(\lambda)
\end{equation}

From HFOSC images, the extinction coefficients in different filters were
determined using Eq. 1. The observed values at $U,B,V,R$ and $I$ 
filters and the theoretical values calculated at their central wavelengths
are given in Table 1. The observed values of extinction  
coefficients have a mean value of 0.36 $\pm$ 0.07,
0.21 $\pm$ 0.04, 0.12 $\pm$ 0.04, 0.09 $\pm$ 0.04 and 0.05 $\pm$ 0.03
in $UBVRI$ bands respectively. A histogram of the measured extinction 
coefficients is
shown in Fig. 1. The monthly 
variation of extinction determined for the period 2000$-$2008 is shown in 
Fig. 2. From Fig. 2 it is seen that the
extinction in summer months is larger than during the winter months.
Here, summer months refer to the period between May$-$September and
winter months refer to the period between October$-$April.
%Extinction in all bands is also found to be clearly correlated with
%the night time average wind speed and this is shown in Fig. 3.
The variation of the measured extinction coefficients
with the average night time wind speed is shown in Fig. 3 for 
all the bands. From Fig. 3, 
there is a hint for a correlation between the extinction coefficient
and the wind speed. But, non-inclusion of one high extinction value
at high wind speed in each band removes the correlation (linear correlation
coefficient R $<$ 0.5) and the linear fit shown in Fig. 3 almost becomes horizontal. Thus, from the present
data set, there is no clear evidence of a correlation between the extinction
coefficient and the average night time wind speed. Further data are needed
to check for the presence or absence of this correlation.
A statistical summary of the measured extinction data is given in Table 2. The 
yearly averages of extinction coefficient in several bands are given in Table 3. 
As evident from Table 3, 
we find no clear evolution of extinction
over the years 2000$-$2008. 

\subsection{Nature of Aerosols at IAO}
Aerosol extinction properties can be studied from the observed total
extinction. 
The observed values of mean extinction were analysed to study
the nature of aerosols at IAO. 
%by substracting the extinction due to Rayleigh scattering and ozone. 
It has been noted by Hayes \& Latham (1975) that the extinction
due to Rayleigh and ozone can be calculated theoretically with an
accuracy of the order of $\pm$0.01 mag/airmass for wavelengths between
3300 $-$ 10800 \AA, using Eqs. 2 and 3. 
%Aerosol extinction is
%represented by the following equation (Hayes \& Latahm 1975)
Therefore, from the total measured extinction,
theoretically calculated values of extinction due to Rayleigh
scattering and ozone were subtracted to get the observed
values of extinction due to aerosols. Thus the extinction due
to aerosols at IAO is estimated as follows

\begin{equation}
A_{aer}(\lambda) = <A> - A_{Ray}(\lambda) - A_{oz}(\lambda) 
\end{equation}

The deduced values of aerosols using Eq. 7 could well be represented
as
\begin{equation}
A_{aer} = \beta \lambda^{-\alpha}
\end{equation}

$\alpha$ and $\beta$ in Eq. 8, were then determined by linear fits on 
the log-log plots of the deduced aerosol extinction and wavelength. 
We find a mean value of $\alpha$ = 0.84 $\pm$ 0.23 and $\beta$ = 0.07 $\pm$ 0.04
from analysis of 14 nights for which a linear fitting was possible.
The mean
value of $\alpha$ found here is similar to that found for many
observatories (Hayes \& Latham 1975).  A comparison of the extinction at
Hanle with that of other sites is given in Table 4. From Table 4 it is 
seen that the extinction at IAO is similar to that of the best astronomical 
sites.
%Also, the scale height being
%a function of elevation of the site, must mean that H is a function
%of atmospheric temperature (Krisciumas 1990).

\begin{figure}[t]
\hspace*{-3.0cm}\psfig{file=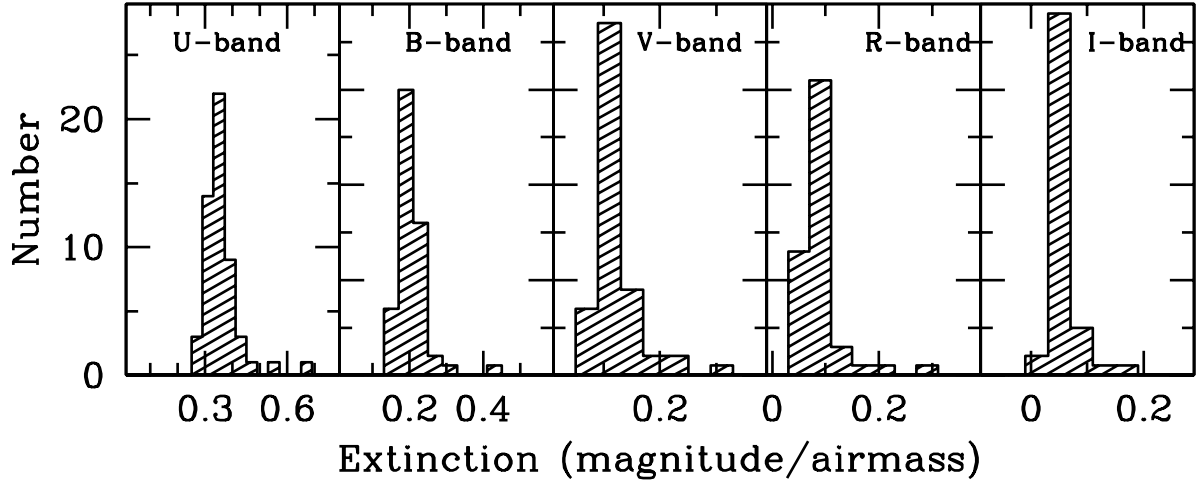}
\vspace*{-2.0cm}
\caption{Distribution of the measured extinction coefficients in magnitude/airmass at IAO in $UBVRI$ bands}
\end{figure}

\begin{figure}[t]
\psfig{file=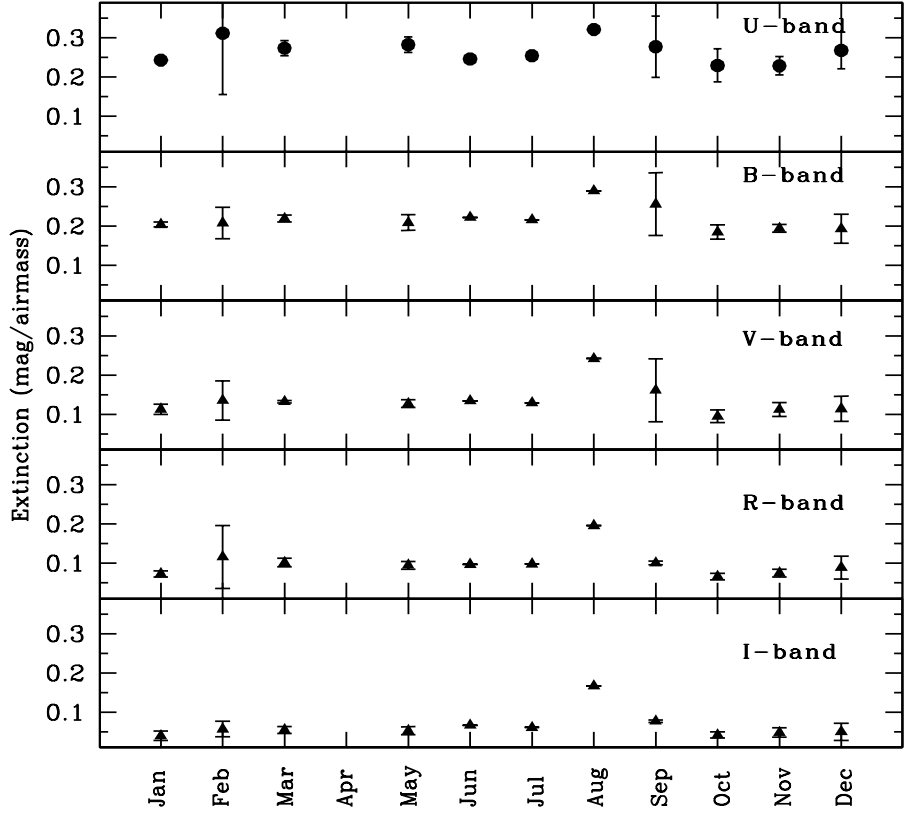,width=14cm,height=13cm}
\caption{Monthly variation of extinction at IAO during the period 2003$-$2007. 
The error bars when not visible are less than the size of  
symbols. These error bars
represent the scatter in the values, the accuracy of individual measurements
being much smaller.}
\end{figure}

\begin{figure}[t]
\psfig{file=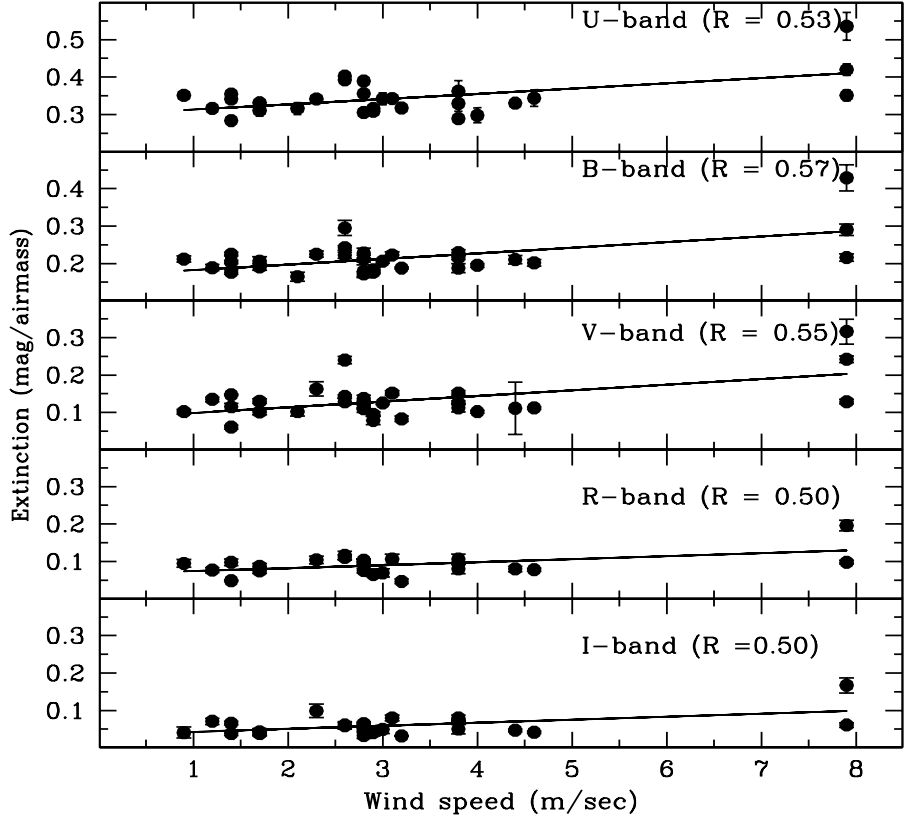,width=14cm,height=13cm}
\caption{Variation of extinction in $UBVRI$ passbands with the average night time wind speed. The 
linear correlation coefficient in each band is given in the respective 
panel. The error bars when not visible are smaller than the symbol size. The solid
line is the unweighted linear least squares fit to the data.} 
\end{figure}

\begin{table}
\centering
\caption{The calculated and measured values of the extinction coefficients in mag/airmass at Hanle}
\begin{tabular} {ccccccc} \hline
Filter & $\lambda_{0\AA}$& $k_{Ray}$&  $k_{aer}$ &  $k_{oz}$ &  $k_{sum}$ & Observed   \\ \hline
U      & 3650 &  0.3307  &  0.0099  & 0.0008  & 0.3414   & 0.36 $\pm$ 0.07 \\
B      & 4400 &  0.1522  &  0.0085  & 0.0005  & 0.1612   & 0.21 $\pm$ 0.04 \\
V      & 5500 &  0.0610  &  0.0071  & 0.0262  & 0.0943   & 0.12 $\pm$ 0.04 \\
R      & 7000 &  0.0229  &  0.0059  & 0.0036  & 0.0324   & 0.09 $\pm$ 0.04 \\
I      & 8800 &  0.0091  &  0.0049  & 0.0000  & 0.0140   & 0.05 $\pm$ 0.03 \\ \hline
\end{tabular}
\end{table}

\begin{table}
\centering
\caption{Extinction coefficients in mag/airmass at Hanle for the whole period as well as for summer and winter months}
\begin{tabular}{ccccccc} \hline
Filter & \multicolumn{2}{c}{Total} & \multicolumn{2}{c}{Summer} & \multicolumn{2}{c}{Winter} \\ 
       & Nights  & Mean   & Nights  & Mean  & Nights  & Mean  \\  \hline
U      & 54  & 0.36 $\pm$ 0.07  & 10 & 0.38 $\pm$ 0.06 & 44 & 0.35 $\pm$ 0.07 \\
B      & 57  & 0.21 $\pm$ 0.04  & 11 & 0.24 $\pm$ 0.07 & 46 & 0.20 $\pm$ 0.03 \\
V      & 58  & 0.12 $\pm$ 0.04  & 11 & 0.15 $\pm$ 0.06 & 47 & 0.12 $\pm$ 0.03 \\
R      & 50  & 0.09 $\pm$ 0.04  & 08 & 0.11 $\pm$ 0.03 & 42 & 0.09 $\pm$ 0.04 \\
I      & 47  & 0.05 $\pm$ 0.03  & 08 & 0.08 $\pm$ 0.04 & 39 & 0.05 $\pm$ 0.02 \\ \hline
\end{tabular}
\end{table}

\begin{table}
\centering
\caption{Average yearly Extinction in mag/airmass  at IAO}
\begin{tabular} {cccccc} \hline
Year & U  & B & V & R & I  \\  \hline
2000 & 0.38 $\pm$ 0.04 & 0.19 $\pm$ 0.03 & 0.12 $\pm$ 0.02 & 0.09 $\pm$ 0.02 & 0.05 $\pm$ 0.01 \\
2002 & 0.35 $\pm$ 0.04 & 0.21 $\pm$ 0.03 & 0.12 $\pm$ 0.04 & 0.09 $\pm$ 0.04 & 0.05 $\pm$ 0.03 \\
2003 & 0.38 $\pm$ 0.08 & 0.25 $\pm$ 0.10 & 0.15 $\pm$ 0.09 & 0.07 $\pm$ 0.01 & 0.03 $\pm$ 0.01 \\ 
2004 & 0.34 $\pm$ 0.06 & 0.21 $\pm$ 0.06 & 0.13 $\pm$ 0.06 & 0.09 $\pm$ 0.04 & 0.06 $\pm$ 0.04 \\
2005 & 0.37 $\pm$ 0.10 & 0.22 $\pm$ 0.03 & 0.14 $\pm$ 0.04 & 0.11 $\pm$ 0.06 & 0.06 $\pm$ 0.02 \\
2006 & 0.34 $\pm$ 0.04 & 0.21 $\pm$ 0.01 & 0.12 $\pm$ 0.01 & 0.09 $\pm$ 0.02 & 0.06 $\pm$ 0.01 \\
2007 &                   &                   & 0.09 $\pm$ 0.00 & 0.06 $\pm$ 0.00 & 0.03 $\pm$ 0.00  \\ 
2008 &                   & 0.19 $\pm$ 0.04   & 0.10 $\pm$ 0.05 & 0.07 $\pm$ 0.02 & 0.06 $\pm$ 0.03 \\ \hline
\end{tabular}
\end{table}

\begin{table}
\centering
\caption{Comparison of extinction in mag/airmass  at various sites}
\begin{tabular}{lrcccccl} \hline
Site      & Altitude & U       & B       & V         &  R    & I     &  Reference \\  
          & (m)      &  &  & & & \\ \hline
Rangapur  &  695     & 0.7-0.9 & 0.4-0.4 & 0.26-0.32 &       &       &  Kulkarni \& Abhyankar 1978 \\
IGO       &  1005    &         & 0.46    & 0.28      &       &       &  Das et al. 1999 \\
Nainital  &  1951    & 0.57    & 0.28    & 0.17      & 0.11  & 0.07  &  Kumar et al. 2000 \\
Devasthal &  2450    & 0.49    & 0.32    & 0.21      & 0.13  & 0.08  &  Mohan et al. 1999 \\
Kavalur   &  725     & 0.75    & 0.34    & 0.23      &       &       &  Singh et al. 1988   \\
Leh       & 3500     & 0.50    & 0.28    & 0.17      &       &       &  Singh et al. 1988  \\
KPNO      & 2120     & 0.622   & 0.281   & 0.162     & 0.119 & 0.075 & Landolt \& Uomoto 2007 \\
La Silla  & 2400     & 0.424   & 0.271   & 0.164     &       &       & Giraud et al. 2006 \\
ALMA      & 5000     & 0.260   & 0.160   & 0.110     &       &       & Giraud et al. 2006 \\
Mauna Kea & 4200     & 0.358   & 0.198   & 0.119     & 0.100 & 0.050 & Mauna Kea 2005 \\
Hanle     & 4467     & 0.36    & 0.21    & 0.12      & 0.09  & 0.05  &  This work       \\  \hline
\end{tabular}
\end{table}

\begin{table}
\centering
\caption{Yearly statistics of night sky brightness in mag/arcsec$^2$ at IAO. The errors given
here are the standard deviation of the yearly measurements. In 2006, $B$ and $I$ bands have only
one measurement each, and the errors are therefore not given}
\begin{tabular}{cccccc}\hline
Year & U  &  B   & V   &  R   &   I  \\ \hline
2003 & 22.19 $\pm$ 0.23 & 22.57 $\pm$ 0.09 & 21.21 $\pm$ 0.46 & 20.48 $\pm$ 0.48 & 18.79 $\pm$ 0.28 \\
2004 & 22.33 $\pm$ 0.18 & 22.49 $\pm$ 0.26 & 21.30 $\pm$ 0.21 & 20.56 $\pm$ 0.38 & 18.90 $\pm$ 0.33 \\
2005 & 22.10 $\pm$ 0.35 & 22.45 $\pm$ 0.22 & 21.32 $\pm$ 0.16 & 20.66 $\pm$ 0.29 & 18.92 $\pm$ 0.40 \\
2006 &                  & 22.30            & 21.21 $\pm$ 0.20 & 20.57 $\pm$ 0.19 & 18.58             \\ 
2007 & 21.84 $\pm$ 0.22 & 22.14 $\pm$ 0.46 & 21.11 $\pm$ 0.20 & 20.34 $\pm$ 0.36 & 18.69 $\pm$ 0.20  \\  \hline
\end{tabular}
\end{table}

\begin{table}
\centering
\caption{Comparison of night sky brightness in mag/arcsec$^2$ at various sites}
\begin{tabular}{lcccccl} \hline
Site  & U & B & V & R & I & Reference \\ \hline
La Silla           & ---  & 22.8 & 21.7 & 20.8 & 19.5 & Mattila et al. 1996 \\
Calar Alto         & 22.2 & 22.6 & 21.5 & 20.6 & 18.7 & Leinert et al. 1995 \\
La Palma           & 22.0 & 22.7 & 21.9 & 21.0 & 20.0 & Benn \& Ellison 1998 \\
KPNO               & ---  & 22.9 & 21.9 &  --- & ---  & Pilachowski et al. 1989 \\
Mt. Graham         & 22.38 & 22.86 & 21.72 & 21.9 & --- & Taylor et al. 2004 \\
CTIO               & 22.12 & 22.82 & 21.79 & 21.19 & 19.85 & Krisciunas et al. 2007 \\
Paranal            & 22.35 & 22.67 & 21.71 & 20.93 & 19.65 & Patat  2008 \\
SPM, Mexico        & 21.50 & 22.30 & 21.40 & 20.70 & 19.20 & Tapia et al. 2007 \\
Hanle              & 22.14 & 22.42 & 21.28 & 20.54 & 18.86 & This work         \\ \hline
\end{tabular}
\end{table}

\begin{figure}[t]
\psfig{file=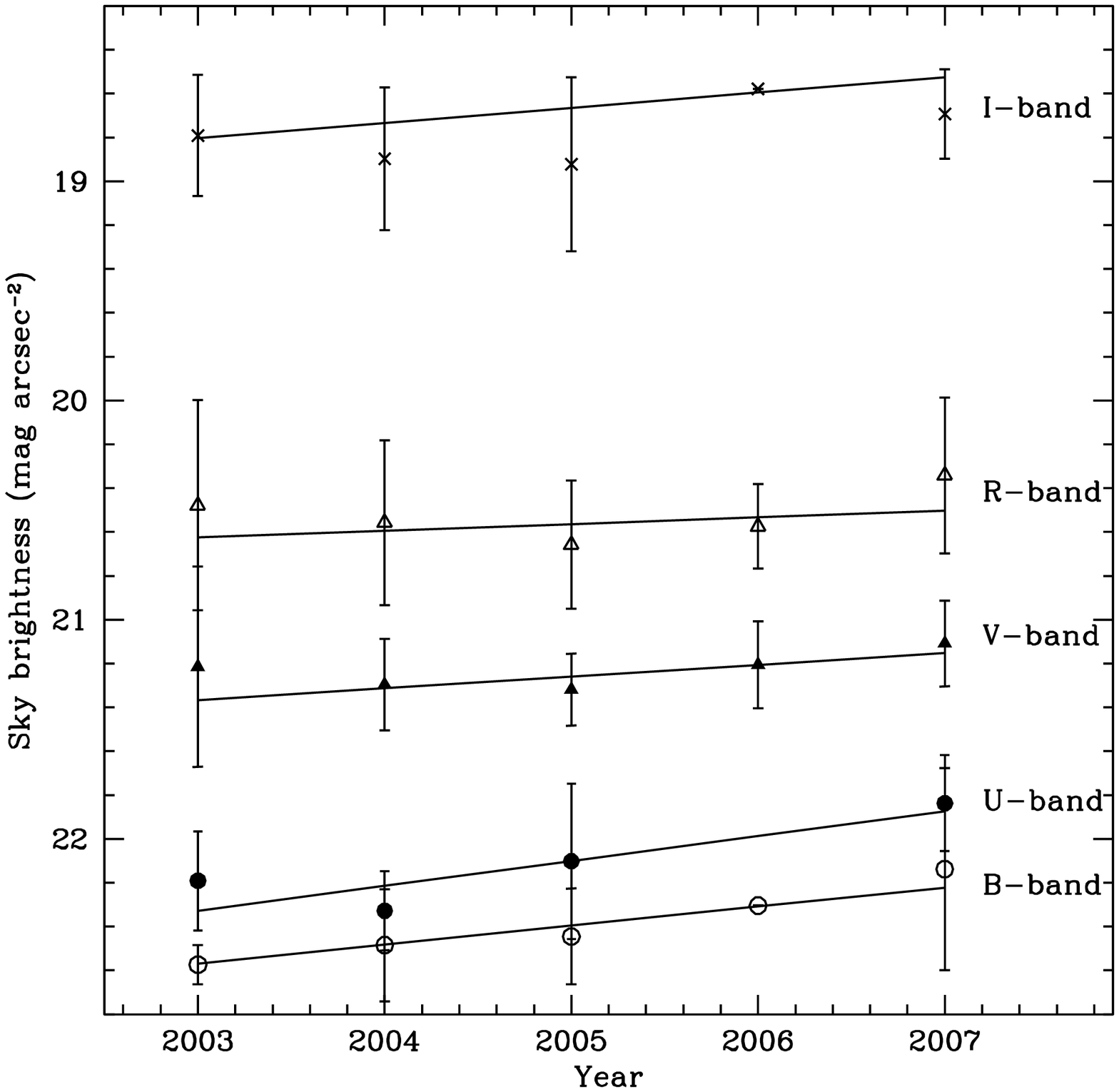,width=13cm,height=12cm}
\caption{Yearly variation of night sky brightness in $UBVRI$ bands. The solid line is 
the unweighted linear least squares fit to the data. The error bars represent the scatter
in the average values, the accuracy of individual measurements being much smaller. For the 
$B$ and $I$ bands in 2006 only one measurement is available in each band and hence
the error bars are not plotted.}
\end{figure}

\section{Sky brightness}
The night sky  brightness is estimated following the relation given by Krisciunas et al. (2007):

\begin{equation}
S = -2.5 log\left (\frac{C_{sky}}{C_\ast}\right) + 2.5log \left(\frac{E_{sky}}{E_\ast}\right) + K_{\lambda}X_{\ast} + M_{\ast}
\end{equation}

where S is the brightness of the sky in magnitudes,  $C_{\ast}$ is the total 
counts above sky 
within the aperture of the 
standard star of magnitude $M_{\ast}$ with an exposure 
time $E_{\ast}$, $C_{sky}$ is the mean
sky counts times the area of the aperture with exposure $E_{sky}$. 
$X_{\ast}$ is the airmass and $k_{\lambda}$ is the atmospheric
extinction corresponding to the filter used.
The sky brightness in magnitude per square arcsec,  I($\mu$) is then 

\begin{equation}
I(\mu) = S + 2.5logA
\end{equation}

Here, A is the area of the aperture in square arcseconds estimated
from the plate scale of the CCD. It is known that the
sky becomes brighter at larger airmass. This is due to the 
natural effect of airglow which is brighter at low elevations.
Light pollution also makes the sky brighter at low elevations.
On the other hand, the contribution of extra-terrestrial 
component to sky brightness is independent of zenith distance.
The images for sky brightness were acquired at various airmass ranges
and therefore, the measured values of sky brightness need to be corrected for
its dependence on zenith distance caused by the effects of 
airglow. To get the sky brightness at zenith, a correction
($\Delta m$) has been added to the measured
sky brightness following Patat (2003).

\begin{equation}
\Delta m = -2.5log_{10}\left[(1-f) + fX\right] + k(X-1)
\end{equation}

This assumes that a fraction($f$) of the total sky brightness
is generated by airglow, and the remaining (1-$f$) fraction
is produced outside the atmosphere (thus including
zodiacal light, faint stars and galaxies). To convert
the measured sky brightness to zenith, the mean extinction
values at IAO and $f$ = 0.6 (Patat 2003) were used. Here
$X$ is the optical path length along a line of sight (not
quite equivalent to the secant of zenith angle) and is given as 
(Patat 2003)

\begin{equation}
X = \left(1 - 0.96sin^{2}Z\right)^{-1/2}
\end{equation}

In order to estimate the night sky brightness during the dark moon
period, the following criteria were applied while choosing the data
from the IAO archives. They are (a)  
%Imaging data extracted from IAO archives for sky brightness 
%estimation, includes supernova observations done under a wide variety of 
%observing conditions. As we are interested in the sky 
%brightness of a completely dark sky, from the original data set, 
%only those observations which satisfy the following conditions of a 
%dark sky were then extracted. They are 
photometric conditions, (b) airmass $\le$ 1.4, (c) galactic 
latitude $\vert$b$\vert$ $>$ 10$^\circ$, (d)  time distance from the 
closest twilight $\Delta t$ $>$ 1 hr and (e) no moon (fractional
illumination of moon equal to zero or moon elevation $<$ $-18^{\circ}$).
After applying these restrictions, we collected 210 frames, taken over
47 nights during 2003$-$2007.
In this study, secondary standard stars (more than 6) were used in each frame and thus 
$E_{\ast} = E_{sky}$ in Eq. 9. 
It should be noted that the sky brightness at any line
of sight towards the sky is not corrected for extinction following the 
method adopted in studies of sky brightness (Krisciunas et al. 2007).

The average moonless night sky brightness at the zenith at IAO are 
22.14 $\pm$ 0.32 in $U$, 22.42 $\pm$ 0.30 in $B$, 
21.28 $\pm$ 0.20 in $V$, 20.54 $\pm$ 0.37 in $R$ and
18.86 $\pm$ 0.35 in $I$ band. Their year wise statistics
is given in Table 5 and their evolution over the year is 
shown in Fig. 4. The sky brightness evolution might not be
significant and considering the errors, this might be
consistent with no evolution. A comparison with other sites is given in Table 6.
The sky brightness values at IAO listed in Table 6 are similar to 
those of other astronomical sites in $UBVR$ except $I$ band.  
The sky at IAO in the $I$ band is clearly brighter than the 
other astronomical sites listed in Table 6 with the exception of Calar Alto. 
This might be due to the hydorxy OH (Meinel) bands being stronger at IAO (see Section 5).

\begin{figure}[t]
\psfig{file=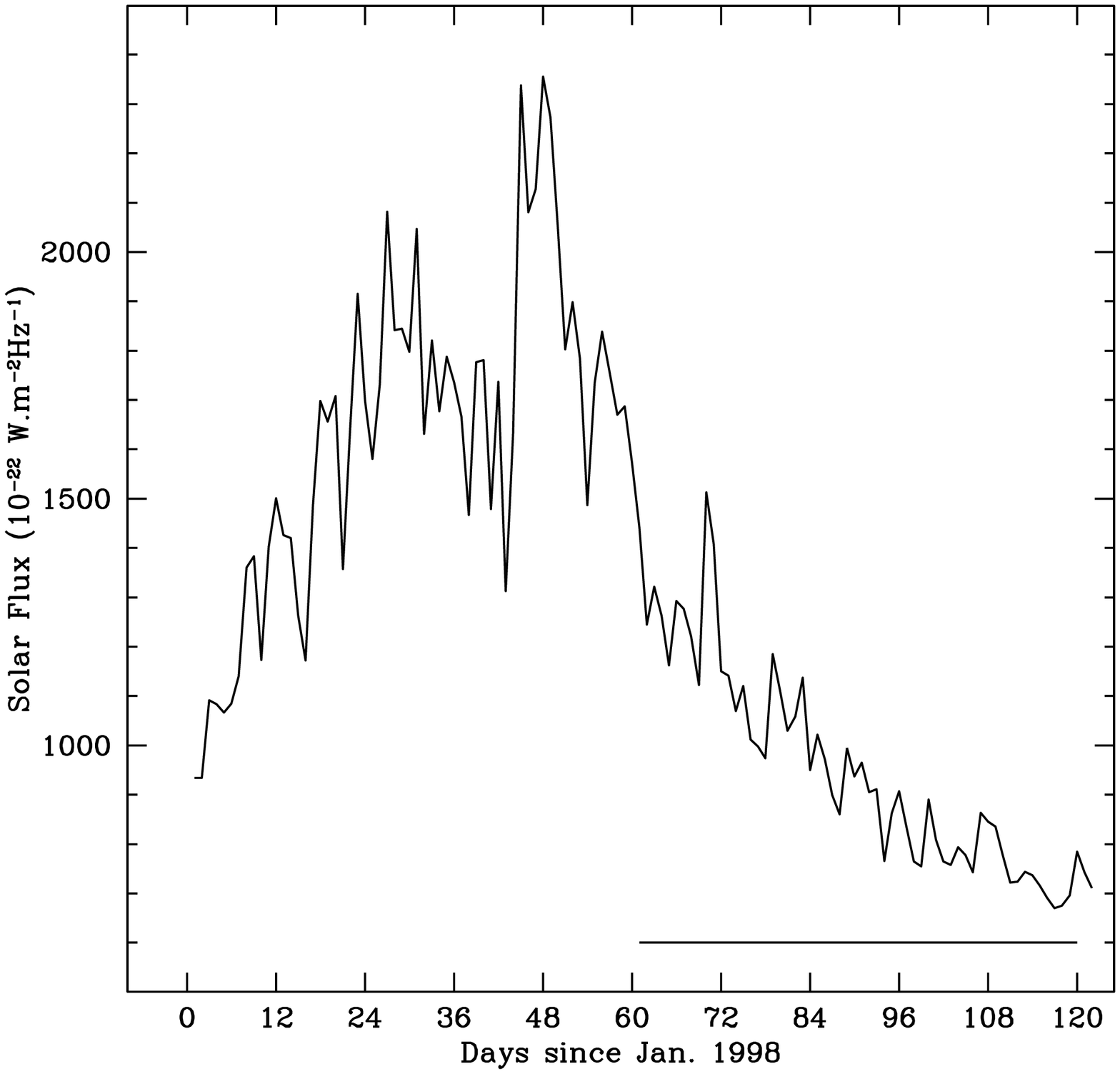,width=13cm,height=12cm}
\caption{Average monthly solar flux at 10.7 cm since January 1998. The 
horizontal bar indicates the period (2003$-$2007) of sky brightness results presented here.}
\end{figure}

\begin{figure}[t]
\psfig{file=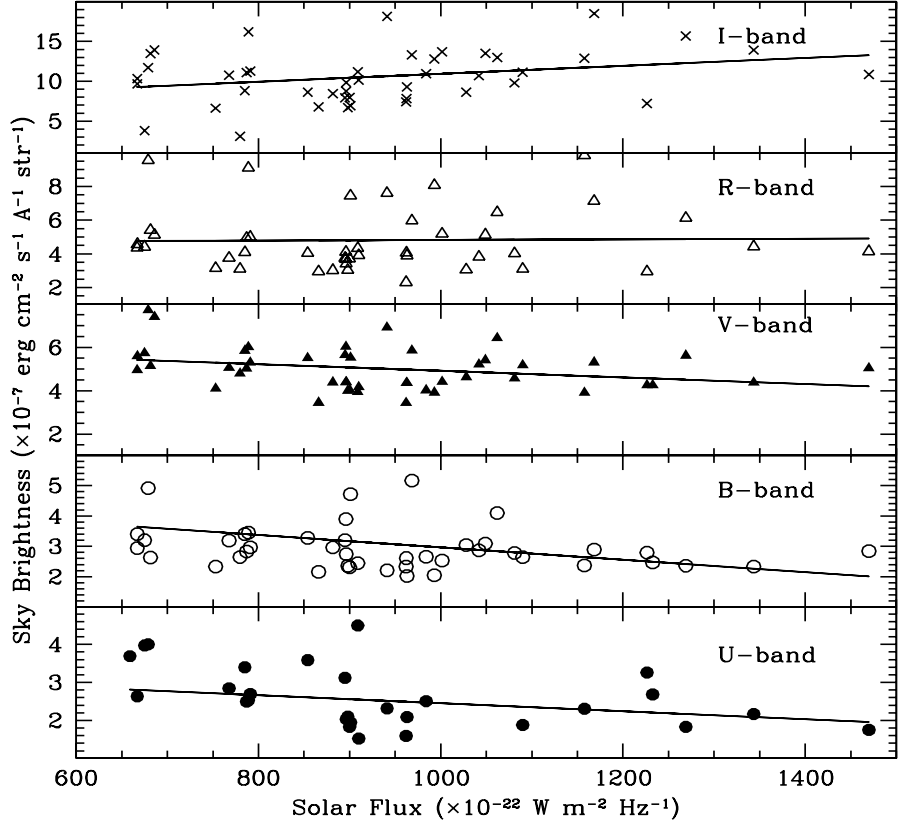,width=13cm,height=12cm}
\caption{Variation of sky brightness with solar flux density at 10.7 cm.
The solar flux density values are 10 days average prior to the date of
measurement of sky brightness. The solid line is the linear least squares
fit to the data.}
\end{figure}

\begin{figure}[t]
\hspace*{-1.2cm}\psfig{file=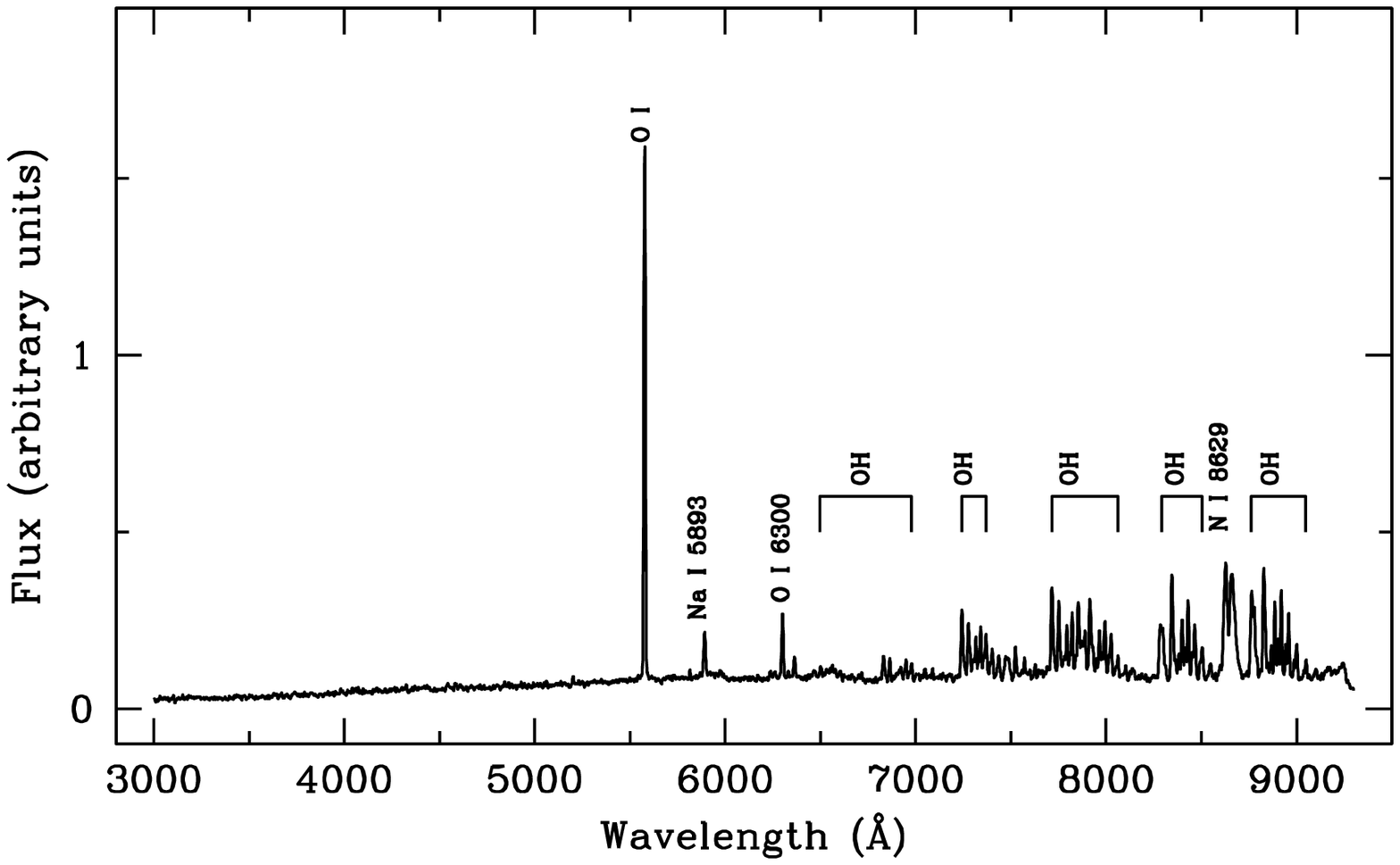,width=16cm}
\caption{Night sky spectrum at IAO, Hanle. Prominent sky lines are marked.}
\end{figure}

\subsection{Variation of night sky brightness with Solar activity}

A possible correlation between sky brightness and solar activity
was first pointed out by Rayleigh (1928) and Rayleigh \& Jones (1935).
It was later confirmed by several authors (Walker 1988; Patat 2008;
Krisciunas et al. 2007). The monthly averaged
2800 MHz solar flux\footnote {http://www.ngdc.noaa.gov/stp/SOLAR/ftpsolarradio.html} between 
Jan 1998$-$March 2008 is shown in Fig. 5.  The data used
in this work for sky brightness measurements covers the period 2003
to 2007 (shown as a horizontal bar in Fig. 5), and thus spans
the declining phase of solar activity cycle$-$23. Our data
can therefore be used to look for possible correlation between sky brightness
and solar activity. For this, it might be more practical to use
linear unit for sky brightness, contrary to the common astronomical
practice of expressing sky brightness in magnitude per square arcsec.
Following Patat (2003) the sky brightness in linear
scale (expressed in $erg s^{-1} cm^{-2} \AA^{-1} sr^{-1}$) is given as

\begin{equation}
B_{\lambda} = 10^{\left(-0.4(m_{sky,\lambda} - m_{0,\lambda} - 26.573) \right)}
\end{equation}

Here, $m_{0,\lambda}$ is the magnitude zero point (Cox 2000), 
and $m_{sky,\lambda}$ is the measured sky brightness
at zenith in mag.arcsec$^{-2}$. Correlations have been looked for between
measured sky brightness and average solar flux computed for 10 days
prior to each sky brightness measurement. The results
are shown in Fig. 6. A trend for a correlation is seen, though 
differently in various filters. To have
a quantitative description of the correlation, a linear fit
of the data to $m = m_0 + \gamma F_{sky}$ was done and the 
results of the fit are given in Table 7. From the low
correlation coefficients, we point out that there is no
correlation between sky brightness and solar activity, in this present 
data set. Similary, no correlation between sky brightness and solar activity is found between the summer and
winter months.

\begin{table}

\centering
\caption {Linear regression analysis of night sky brightness v/s solar 
activity; $I_{sky} = a. Flux_{star} + b$}
\begin{tabular} {ccccc} \hline
Filter & a & b & R & npts \\ \hline
U      & 3.51 $\times 10^{-07}$ $\pm$ 7.97 $\times 10^{-08}$  & -1.05 $\times 10^{-10}$ $\pm$ 7.85 $\times 10^{-11}$  & -0.28  & 28      \\
B      & 5.01 $\times 10^{-07}$ $\pm$ 9.06 $\times 10^{-08}$  & -2.04 $\times 10^{-10}$ $\pm$ 9.47 $\times 10^{-11}$  & -0.31  & 45      \\
V      & 6.44 $\times 10^{-07}$ $\pm$ 7.11 $\times 10^{-08}$  & -1.52 $\times 10^{-10}$ $\pm$ 9.47 $\times 10^{-11}$  & -0.29  & 47      \\
R      & 4.64 $\times 10^{-07}$ $\pm$ 1.45 $\times 10^{-07}$  &  1.86 $\times 10^{-11}$ $\pm$ 1.53 $\times 10^{-10}$  &  0.02  & 45      \\
I      & 5.93 $\times 10^{-07}$ $\pm$ 2.52 $\times 10^{-07}$  &  5.00 $\times 10^{-10}$ $\pm$ 2.66 $\times 10^{-10}$  &  0.28  & 45       \\ \hline
\end{tabular}
\end{table}

\section{Night sky spectrum}
A typical night sky spectrum is shown in Fig. 7, and the data acquired for
generating the spectrum is described in Section 2. The strong 
emission lines/bands clearly identified in the spectrum have been labelled with
their corresponding atomic/molecular names, and a complete list of all the lines/bands
identified in the spectra along with their respective strengths are
listed in Table 8. The distinctive features seen in 
the night sky spectrum are the OI lines (5577 and 6300 \AA),   
OH rotational vibrational Meinel bands in the red region of 
the spectrum and lines due to Na and N. 
%Apart from the airglow features, lines due to 
%atomic species of Na and N are also seen in the spectrum.

\begin{table}
\centering
\caption{Identified lines/bands in the night sky spectrum at IAO}
\begin{tabular}{ccr|ccr} \hline
Wavelength  &  Species & E.Width  & Wavelength & Species & E. Width\\ 
 (\AA)      &          &  (\AA)   & (\AA)      &         & (\AA)\\ \hline
  5577.3    & [OI]     & 154.4   & 8062.4    &  OH      &   3.9  \\
  5891.6    & Na I     &  17.3   & 8290.6    &  OH      &  56.3  \\
  6237.4    & OH       &   4.2   & 8344.6    &  OH      &  30.9  \\
  6261.4    & OH       &   4.2   &  8430.3   &  OH      &  18.2  \\
  6300.3    & [OI]     &  23.1   &  8452.2   &  OH      &   6.8  \\
  6329.9    & OH       &   1.7   &  8465.5   &  OH      &  18.2 \\
  6363.8    & [OI]     &   8.5   &  8504.6   &  OH      &  13.4 \\
  6498.7    & OH       &   4.7   &  8629.2   &  N I     &  50.9 \\
  6832.6    & OH       &  10.1   &  8655.9   &  N I     &  72.3 \\
  6865.7    & OH       &  10.9   &  8763.7   &  OH      &  45.0 \\
  6978.6    & OH       &   4.4   &  8778.3   &  OH      &  25.0 \\
  7242.2    & OH       &  30.3   &  8790.9   &  OH      &   9.0 \\
  7275.0    & OH       &  20.8   &  8827.1   &  OH      &  37.1 \\
  7316.2    & OH       &   7.3   &  8867.6   &  OH      &   8.6 \\
  7341.0    & OH       &  14.8   &  8885.8   &  OH      &  17.6 \\
  7369.2    & OH       &  10.5   &  8903.1   &  OH      &   7.7 \\
  7715.8    & OH       &  26.6   &  8919.7   &  OH      &  20.9 \\
  7750.7    & OH       &  20.5   &  8958.2   &  OH      &  15.8 \\
  7791.1    & OH       &   8.0   &  8987.6   &  OH      &   5.5 \\
  7821.6    & OH       &  10.3   &  9001.1   &  OH      &  13.4 \\
  7853.6    & OH       &   9.6   &  9049.8   &  OH      &   6.7 \\
 \hline

\end{tabular}
\end{table}

\section{Conclusion}
We have studied the nature of the night sky at IAO. The results are
summarized below:
\begin{enumerate}
\item The measured average extinction coefficient at IAO during the period 2003$-$2008
are 0.36 $\pm$ 0.07 in $U$, 0.21 $\pm$ 0.04 in $B$, 0.12 $\pm$ 0.04 
in $V$, 0.09 $\pm$ 0.04 in $R$ and 0.05 $\pm$ 0.03 in $I$. However, 
the average extinction during summer months is slightly larger than
that of winter months. There is no clear evidence for a correlation
between the measured extinction coefficient and the average night time
wind speed.
\item The moonless night sky brightness at zenith are 22.14 $\pm$ 0.32
in $U$, 22.42 $\pm$ 0.30 in $B$, 21.28 $\pm$ 0.20 in $V$, 20.54 $\pm$ 0.37
in $R$ and 18.86 $\pm$ 0.35 in $I$. Except for the $I$ band, the sky brightness
in other bands at IAO are similar to those of other dark sites in the world. 
The bright
nature of the sky in $I$ band at IAO, might be due to the presence of strong
OH rotation vibrational Meinel bands in the red region of the optical 
spectrum.  We find no dependence of the night sky brightness
with 10.7 cm solar flux, probably due to insufficient data coverage 
during the solar cycle. 
%This might be due to the reason that the data
%presented here covers only a small portion of the solar activity cycle, that
%too during the declining phase of solar cycle-23. Further brightness
%measurements covering a complete solar cycle is therefore needed, 
%to know the correlation between sky brightness and solar activity if any at IAO.
\item The moonless night sky spectrum covering the wavelength
range 3000 to 9300 has been presented. Features from OI, OH, Na and N are 
seen in the spectra. 
Lines due to light pollution
are not seen in the spectrum, and thus Hanle is free from any
man made light pollution.
\end{enumerate}

We conclude that IAO has a night sky similar to the best sites in the world. 
In addition to the good skies, IAO also has a 
longitudinal advantage in covering the longitudinal gap between
observatories in the West and the East.
%comparison with other major observatories
%in the globe. 
Hanle region could thus be a potential site for any future
large Indian optical-infrared telescope(s).

\section{Acknowledgement}
The authors thank the anonymous referee for his critical comments.
The help rendered by the large number of individuals from  IIA in the
site characterisation of Hanle is thankfully acknowledged.

\end{document}